\documentclass[11pt]{article}

\usepackage[margin=1in]{geometry}

\usepackage{amsmath,amssymb,amsfonts}

\usepackage{moreverb,url}
\usepackage[colorlinks,bookmarksopen,bookmarksnumbered,citecolor=blue,urlcolor=blue]{hyperref}
\usepackage{soul}
\usepackage{subfig}
\usepackage{tikz}
\usetikzlibrary{bayesnet}
\usepackage{bm}
\usepackage{graphicx}
\usepackage{algorithm}
\usepackage{natbib}
\usepackage{booktabs}
\usepackage{xcolor}

\newcommand{\E}{\mathbb{E}}
\newcommand{\bX}{\bm{X}} 

\newcommand{\bW}{\bm{W}}  

\begin{document}

\title{Detecting and Mitigating Treatment Leakage in Text-Based Causal Inference: Distillation and Sensitivity Analysis}

\author{
  Adel Daoud\thanks{Corresponding author. Email: adel.daoud@liu.se} \\
  Institute for Analytical Sociology, Link\"oping University, Sweden \\
  Department of Computer Science and Engineering, Chalmers University of Technology, Sweden \\
  \and
  Richard Johansson \\
  Department of Computer Science and Engineering, Chalmers University of Technology, Sweden \\
  \and
  Connor Jerzak \\
  Department of Government, University of Texas at Austin, USA
}

\date{}

\maketitle

\begin{abstract}
Text-based causal inference increasingly employs textual data as proxies for unobserved confounders, yet this approach introduces a previously undertheorized source of bias: treatment leakage. Treatment leakage occurs when text intended to capture confounding information also contains signals predictive of treatment status, thereby inducing post-treatment bias in causal estimates. Critically, this problem can arise even when documents precede treatment assignment, as authors may employ future-referencing language that anticipates subsequent interventions. Despite growing recognition of this issue, no systematic methods exist for identifying and mitigating treatment leakage in text-as-confounder applications. This paper addresses this gap through three contributions. First, we provide formal statistical and set-theoretic definitions of treatment leakage that clarify when and why bias occurs. Second, we propose four text distillation methods---similarity-based passage removal, distant supervision classification, salient feature removal, and iterative nullspace projection---designed to eliminate treatment-predictive content while preserving confounder information. Third, we validate these methods through simulations using synthetic text and an empirical application examining International Monetary Fund structural adjustment programs and child mortality. Our findings indicate that moderate distillation optimally balances bias reduction against confounder retention, whereas overly stringent approaches degrade estimate precision.
\end{abstract}


\bigskip

\section{Introduction}\label{s:intro}
Given the exponential growth of digital data sources, social scientists increasingly use text for causal inference in observational studies \citep{roberts2020}. While scholars previously relied mainly on surveys or other structured tabular data to measure confounding, they can now mobilize unstructured text data, call it $\bW$, from digitized-archive documents, social media posts, policy statements, medical records, and similar sources. By \textit{structured tabular} data, we mean information that has been systematically processed and annotated by a human to measure a phenomenon of interest (i.e., treatments, outcomes, and confounders) and made readily available in a tabular data matrix for causal estimation; conversely, \textit{unstructured} data is information that lacks such human processing and annotation. A confounder is a common cause of both the treatment (a policy, action, or exposure) and the outcome of interest. Controlling for all confounders is critical to estimating an unbiased causal effect ($\tau$) of a treatment on an outcome. Although tabular data tend to capture many confounders across a variety of research settings, unstructured text data provide an increasingly critical complementary source. Recent methodological frameworks integrate unstructured text data into causal estimation \citep{mozer2020}, often incorporating text without further human processing and instead relying on an algorithmic representation (e.g., via a topic model).

However, this incorporation assumes that an unstructured text source contains information only about the confounder of interest, not about a unit's treatment status. This \emph{no-treatment leakage} assumption (or \emph{measuring purely confounding}) \citep{daoud2022conceptualizing} requires that the treatment variable does not affect---or leak information into---the production of that text. If treatment leakage exists, then using the text to adjust for confounding will lead to post-treatment (collider) bias, because the text also contains information about the treatment. In this sense, treatment leakage is an instance of conditioning on a ``bad control''---a variable (here, in text) that is causally downstream of treatment---so adjustment can open spurious paths and bias causal estimates \citep{cinelli2024crash}.

To mitigate treatment leakage, the text requires distillation before usage \citep{daoud2022conceptualizing}; scholars need to distill (separate) the part of the text that contains information about the confounder (the desired part) from the text that contains information affected by the treatment (the undesired part). Conducting such distillation manually across a large corpus is costly and perhaps unfeasible when the text reveals sentiments about treatment and confounding status (e.g., emotional reactions to a medicine or social policy).

Treatment leakage is a form of post-treatment bias, yet it is not fully reducible to it. Scholars typically assume that collecting confounding data before treatment assignment---pre-treatment data---precludes post-treatment bias \citep{Pearl2015}. One would expect this logic to extend to measuring confounding in text: treatment leakage could not exist if the text was collected before treatment assignment. However, human language behaves differently. Treatment leakage can occur even when the text document was produced before treatment assignment \citep{daoud2022conceptualizing}. Because text documents supply rich information and human language can reference future events, treatment leakage can occur even when text is written \textit{before} treatment. This creates an apparent paradox: unstructured text data produced before treatment may still lead to post-treatment bias when used for confounding adjustment. How is this possible?


This article explains how this paradox arises, how to resolve it in principle, and proposes a toolbox of applied methods (text distillers) to handle treatment leakage. By \textit{handling} leakage, we refer to reducing or evaluating leakage influence in two main scenarios. In the first scenario, a scholar knows that leakage exists with strong magnitude. The text $W$ then requires distillation; without distillation, the study will produce a biased estimate of $\tau$. With more information about how leakage occurs, scholars can impose the required assumptions, structure the leakage problem accordingly, and surgically remove the leaked text parts---either through human annotation or algorithmic assistance. In the second scenario---perhaps more common---leakage is suspected or its strength is unknown. Here, investigators should conduct \textit{a treatment-leakage sensitivity analysis}, running various text distillers that yield different representations of the original text.

A key contribution of this article is demonstrating how to conduct text distillation and treatment-leakage sensitivity analysis. By identifying the mechanics of treatment leakage, scholars developing text-based causal inference methods can better tailor their frameworks to reduce bias in causal estimates, $\tau$. Applied scholars can better calibrate their data-collection procedures to minimize treatment leakage at the research outset or conduct treatment-leakage sensitivity analysis.

Treatment-leakage sensitivity analysis relies on text distillers. A \textit{text distiller} is a function $\phi(\cdot)$ that takes text $\bW$ as input and finds a representation $\phi(\bW)$ that purges treatment leakage from the original text while retaining sufficient information about the confounder $U$. A perfect distiller produces a new text representation $\phi(\bW)=\bW_U$ usable as a proxy for unobserved confounding $U$, free from treatment leakage. Most distillers will not be perfect, but will at least reduce leakage influence. We developed four types of distillers suitable for different applied settings, discussed in the next section.

Using \textit{in silico} (simulation) experiments, this article shows how treatment leakage affects estimation of the causal estimand $\tau$. We formalize under what conditions the causal estimand is identified in the presence of treatment leakage and when text distillation is perfect or near-perfect. Existing studies on treatment leakage use only manual human annotation to remedy leakage \citep{daoud2022conceptualizing}---requiring scholars to process the entire text by hand, making the approach suitable only for small text sizes. We propose several automatic and semi-automatic text distillers based on machine learning (ML) for natural language processing (NLP). These ML-NLP methods can process large amounts of data efficiently, making them better suited for research relying on large datasets. Offering a variety of distillers benefits scholars who must make different assumptions given their research contexts.

To demonstrate treatment-leakage sensitivity analysis mechanics, our study mimics an applied research pipeline. We generate human-readable text, $\bW$, using a GPT-2 model \citep{wooddoughty2021}. This text represents a typical document containing information about treatment status $T$, unobserved (hidden) confounding $U$, and residual text $R$ (text that does not affect $T$, $U$, or $Y$ but may still matter for statistical-estimation consistency). Because we retain information about which paragraphs represent $T$, $U$, or $R$, we can conduct controlled experiments, applying different text distillers on $\bW$ and evaluating how these distillers affect the estimated bias $\hat{\tau}$ targeting the pre-defined causal estimand $\tau$. We analyze how estimates $\hat{\tau}$ become more or less biased when applying text distillation at the paragraph level.

Beyond \textit{in silico} experiments, we demonstrate our methods in an applied study evaluating the causal effects of International Monetary Fund (IMF) policies on child health. We selected this case for two reasons. First, this case has attracted interest from many policymakers \citep{unicef2019} and scholars across several social science disciplines \citep{stuckler2008,dreher2009,daoudImpactInternationalMonetary2017,daoudReinsberg2018,stubbs2020}. Second, causal effect identification in IMF research remains disputed \citep{dreher2009,stubbs2020}, and alternative identification strategies are emerging, including text-based approaches \citep{daoud2019international}. Although such strategies are promising, they likely face treatment leakage threats and require treatment-leakage sensitivity analysis. By conducting such an analysis under certain assumptions, we can empirically and systematically evaluate leakage likelihood and purge it. We show how to conduct this analysis for the IMF case, spell out key general assumptions, and suggest rules of thumb for tracing leakage magnitude.

\section{Treatment Leakage in Text Data}

\subsection{Background: Using Text as a Proxy for an Unobserved Confounder}

\begin{figure}[t]
    \begin{center}
    \begin{tikzpicture}[line width=0.25mm, scale=1.0, transform shape]
        \node[obs] (X) {$\bX$} ; %
        \node[latent, below=of X] (U) {$U$} ; %
        \node[obs, right=of X, xshift=-6mm, yshift=-8.5mm] (T) {$T$} ; %
        \node[obs, right=of T, xshift=-2mm] (Y) {$Y$} ; %
        \node[latent, right=of Y, xshift=-3mm, yshift=0mm] (R) {$R$} ; %
        \node[obs, right=of U, xshift=9.5mm] (W) {$\bW$} ; %
        \edge[-latex] {X,U} {T} ; %
        \edge[-latex] {X,U,T} {Y} ; %
        \edge[-latex] {U} {W} ; %
        \edge[-latex] {R} {W} ; %
    \end{tikzpicture}
    \end{center}
    \caption{The basic observational system we consider in this paper expressed as a causal DAG. It consists of observed variables (shaded): confounders ($\bX$), treatment ($T$), outcome ($Y$), document ($\bW$), and unobserved variables (unshaded): confounder ($U$) and residual factors ($R$). \label{fig:baselinecausalmodel}}
\end{figure}
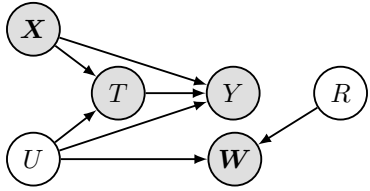

Before addressing treatment leakage, consider the setup expressed as a causal DAG in Figure~\ref{fig:baselinecausalmodel}, where we use observational data to estimate the average treatment effect $\tau$ of a treatment $T$ on an outcome variable $Y$. Valid causal estimates require controlling for confounders. While some confounders are observed ($\bX$), others are unobserved ($U$). Omitting $U$ when estimating $\tau$ may substantially bias the result.

Assume each data point is associated with a document $\bm{W}$ that expresses textual information about the unobserved confounders $U$. Can we use $\bW$ instead of $U$ when estimating $\tau$?
Several methods address this situation \citep{keith2020,mozer2020}, and recent deep learning advances have expanded the causal inference toolkit for complex data types, including text, images, and satellite imagery \citep{balgi2025deeplearning,jerzakImageEconometrics2023,daoudSatelliteIndia2023}.
We make no assumptions about $U$ except that it is unobserved confounding expressed by the text. If we have a more precise idea of what $U$ is, we can potentially use supervised models or zero-shot LLMs to extract a noisy version of $U$ from $\bW$; estimates of $\tau$ based on these noisy values will likely be biased and require correction \citep{egami2023dsl}.

For practical purposes, we assume a \emph{text representation} (or \emph{embedding}) function $\phi$ encodes the text as a high-dimensional numerical variable. Various representation functions exist: bag-of-words representations, topic proportions from Latent Dirichlet Allocation \citep{blei2003}, aggregates over word embeddings \citep{mozer2020}, or Transformer-based text representations \citep{veitch2020}.
The key requirement is that $\phi(\bW)$ preserves information about $U$ expressed by $\bW$ in a format convenient for machine learning and causal estimation methods.

Various methods can estimate the ATE $\tau$ from observational data.
One common method is Inverse Propensity Weighting (IPW, \citealt{rosenbaum1983}), where observations are weighted by the inverse of the \emph{propensity}, or the probability of being treated, $\pi_i={\textrm{Pr}}(T_i=1|\bX_i, U_i)$:
\[\widehat{\tau}_{\mathrm{IPW}} =  \frac{1}{n}\sum_{i=1}^n \left\{ \frac{T_i Y_i}{\pi_i} -\frac{(1-T_i)Y_i}{1-\pi_i}\right\}.\]
Since the exact treatment probabilities $\pi_i$ are unknown, they must be estimated using a \emph{propensity model} that accounts for $\bX$ and $U$. Because $U$ is unobserved, we use $\bW$ (or more precisely, its representation $\phi(\bW)$) instead. The text-based propensity model takes the form
\[
\hat{\pi}_i = \widehat{\Pr}(T|\bX, \phi(\bW))
\]
which can be estimated by probabilistic ML methods such as logistic regression or softmax-based neural networks. Several improvements over basic IPW exist, including doubly robust methods \citep{funk2011doubly}. Combining machine learning for prediction with statistical methods for causal inference represents the ``hybrid modeling culture,'' where predictive algorithms enhance rather than replace inferential goals \citep{daoudDubhashiThreeCultures2023,daoudMeltingPredictionInference2021}.
While we consider only IPW in this article, several text-based causal inference methods have been proposed, including matching \citep{mozer2020}.

The underlying assumptions for text-based IPW deserve emphasis. Crucially, the approach assumes that information about $U$ is preserved: expressed in the text $\bW$ and preserved by the representation $\phi(\bW)$. Given a sufficiently expressive propensity model and adequate training data, IPW estimates using $\bW$ will be equivalent to those obtained if $U$ were observed, recovering the true ATE. 
If this information preservation assumption is violated and the effect of $U$ on text and its representation is weak, the estimated ATE will approach the uncorrected estimate considering only $\bX$ without $U$.


\subsection{Treatment Leakage and why it Matters}
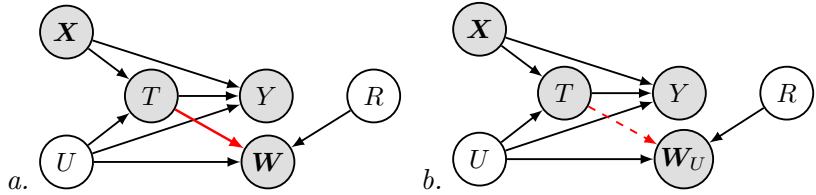
\begin{figure}[htb]
    \begin{center}
\subfloat{\emph{a.}}{
    \begin{tikzpicture}[line width=0.25mm, scale=1.0, transform shape]
        \node[obs] (X) {$\bX$} ; %
        \node[latent, below=of X] (U) {$U$} ; %
        \node[obs, right=of X, xshift=-6mm, yshift=-8.5mm] (T) {$T$} ; %
        \node[obs, right=of T, xshift=-2mm] (Y) {$Y$} ; %
        \node[latent, right=of Y, xshift=-3mm, yshift=0mm] (R) {$R$} ; %
        \node[obs, right=of U, xshift=9.5mm] (W) {$\bW$} ; %
        \edge[-latex] {X,U} {T} ; %
        \edge[-latex] {X,U,T} {Y} ; %
        \edge[-latex] {U} {W} ; %
        \edge[-latex] {R} {W} ; %
        \edge[-latex, line width=0.33mm, red] {T} {W} ; %
    \end{tikzpicture}
}
\subfloat{\emph{b.}}{
    \begin{tikzpicture}[line width=0.25mm, scale=1.0, transform shape]
        \node[obs] (X) {$\bX$} ; %
        \node[latent, below=of X] (U) {$U$} ; %
        \node[obs, right=of X, xshift=-6mm, yshift=-8.5mm] (T) {$T$} ; %
        \node[obs, right=of T, xshift=-2mm] (Y) {$Y$} ; %
        \node[latent, right=of Y, xshift=-3mm, yshift=0mm] (R) {$R$} ; %
        \node[obs, right=of U, xshift=9.5mm] (W) {$\bW_U$} ; %
        \edge[-latex] {X,U} {T} ; %
        \edge[-latex] {X,U,T} {Y} ; %
        \edge[-latex] {U} {W} ; %
        \edge[-latex] {R} {W} ;
        \edge[-latex, dashed, red] {T} {W} ; %
    \end{tikzpicture}
}
\end{center}
\caption{A causal model consisting of observed variables (shaded): confounders ($\bX$), treatment ($T$),  outcome ($Y$), document ($\bW$), and unobserved variables (unshaded):  confounder ($U$) and residual factors ($R$). The red-colored edge in \emph{a.} represents treatment leakage. In \emph{b.}, A distillation function $f$ has removed the treatment information in the text, leaving only information from the confounder. A perfect distiller, $f$, is a procedure that removes all information flowing via the red arrow. When the distiller is applied to ($\bW$), that is equivalent to deleting the red arrow; a less-than-perfect intervention reduces at least the strength of the red arrow. \label{fig:causalmodel}}
\end{figure}

Any causal analysis using text must address treatment leakage, especially when text $\bW$ serves as a proxy for unmeasured confounding. Social scientists increasingly use text as a proxy for unobserved confounders \citep{miao2018identifying,tchetgen2020}. A proxy is a random variable (or vector of random variables) that stands in for another variable, usually an unobserved confounder. Such proxies often imperfectly represent the original variable, yet using them reduces bias. Figure \ref{fig:causalmodel} shows how a proxy works in a directed acyclical graph (DAG), which encodes causal relationships between random variables in an assumed causal system. 
A node represents a random variable or collection of variables; a directed arrow connecting two nodes indicates that the first variable causes the second, with causality flowing in the arrow's direction. Confounding variables $U$ and $X$ are common causes of the causal relationship of interest: the effect of $T$ on $Y$. We denote the average of this effect as $\tau$. In observational studies, leaving a confounder unadjusted leads to bias in the estimator $\hat{\tau}$. In Figure~\ref{fig:baselinecausalmodel}, confounder $X$ is observed but $U$ is unobserved, so the causal effect $\tau$ is not identified because scholars cannot measure $U$ directly.

The next best alternative is measuring a proxy of $U$ to approximate the causal effect. A \textit{proxy} is an imperfect representation of the variable of interest; the more the proxy resembles $U$, the better the proxy and the less bias in the causal estimate. In Figure~\ref{fig:causalmodel}, panel \textit{a}, because $U$ affects text generation in $W$, we can use $W$ as a proxy for $U$. However, $T$ also affects $W$ generation. For example, lawmakers can produce a policy document ($W$) describing the context ($U$) for why they will enforce a new policy $T$. When they do so in the same document, the document encodes information from two DAG paths: first, $U \rightarrow W$, and second, $T \rightarrow W$ (the red arrow). The first arrow represents what scholars want for a good proxy: the stronger this arrow, the better $W$ proxies $U$. The second (red) arrow is undesired, as it injects treatment leakage.

This second arrow clarifies why treatment leakage is a form of post-treatment bias. In the causal flow (reading the DAG left to right), the text proxy $W$ occurs after treatment.

Despite the fact that treatment leakage works like post-treatment bias statistically, treatment leakage is conceptually irreducible to this form of bias. As human language is grammatically flexible, it can reveal future and past intentions, actions, and contexts. Also, the text is written over a longer time period, enabling it to capture more events than a variable in tabular data does. Thus, a text document can reveal information about several events simultaneously, including the treatment assignment, even before this assignment has actually manifested. Therefore, equalizing treatment leakage with post-treatment is not a fully accurate description of this type of leakage.

Consider a thought experiment with two scenarios. In the first scenario, a medical doctor, before she has decided on what treatment to assign (say on January 1), meets a patient \textit{i} and evaluates the patient's health status $U_i$, which represents a key common cause of what treatment the patient should get and the desired outcome. The doctor then writes in the patient's journal, "I will give patient \textit{i}  treatment $T_i=1$, because of $U_i=1$", and thus, the journal $W_i$ is written before the treatment is actually assigned. And then, on January 2, the doctor supplies the patient treatment $T_i=1$ with absolute certainty.

In the second scenario, the doctor still meets with the same patient \textit{i} and evaluates $U_i$ exactly as she did in the first scenario. But the doctor does not write anything in the journal on January 1 before the treatment assignment. She just takes a mental note to give the patient the treatment the next day. On January 2, the doctor does indeed give the patient $T_i=1$. Then on January 3, the doctor updated the patient's journal, writing, "I gave patient \textit{i} the treatment $T_i=1$, because of $U_i=1$".

The difference between the two scenarios is that the first document references the treatment-assignment process in the future, and the second document references the same process in the past. In other words, the only difference in wording is that the first document uses the future grammatical tense of "will give," while the second document uses the past tense "gave." Nonetheless, from a causal identification perspective, the two scenarios are equivalent. Both documents are equally good proxies of $U_i=1$, yet both exhibit the exact same amount of treatment leakage. They both reveal that $T_i=1$ for unit $i$, and thus, using either document will lead to post-treatment bias, despite the fact that the first document was produced before the treatment had been actually assigned.

As discussed in Section \ref{sec:principles}, text distillation offers a way to still use $\bW$, regardless of whether the document was produced pre- or post-treatment. Figure~\ref{fig:causalmodel}, panel \textit{b}, shows what a successful and complete distillation amounts to in a DAG: a distillation that entirely removes leakage in the text, and thereby, the red arrow has been nullified. This nullification is equivalent to applying a text distiller $\phi$ that processes $\bW$ in such a way that a new text product is produced $\bW_U$, free of information regarding $T$. The input of distillation is $\bW$; as the output of the distillation is $\bW_U$, we can remove the red arrow. That is, distillation is like conducting surgery on the DAG.

In practice, however, when the text is complex, and information about $U$ and $T$ is entangled in $\bW$, the best one can hope for is to at least mitigate treatment leakage. In our DAG, this means that the strength of the red arrows has been reduced through distillation.

\section{Developing Text Distillers for Causal Inference}
\label{sec:principles}


\subsection{Two Principles of Text Distillation}
When faced with a treatment-leakage contaminated document $\bW$, the goal of text distillation is to produce a \emph{representation}  $\bm{W'}_U$ free from treatment-leakage traces due to $\bW_T$. As previously stated, a leakage-free document is a document that formally fulfils the property of conditional independence: $\bm{W'}_U$ is conditionally independent of $T$ given $U$. However, such independence on its own is not enough for a representation to be practically useful for causal inference since any trivial representation (e.g., a constant or completely random representation) would satisfy this property. So for a perfect distillation, we also need the requirement that $\bm{W'}_U$ must preserve all the information about $U$ that was originally present in $\bW_U$ when composing the complete document $\bm{W'}$. Whatever level of faithfulness $\bW_U$ had in representing $U$, that is also the level of representativeness $\bm{W'}_U$ must preserve.


We say that a representation $\bm{W'}_U = \phi(\bW)$ is a \emph{perfect distillation} of $\bm{W}_U$ from $\bm{W}$ if the following properties hold:

\begin{enumerate}
\item $\bm{W'}_U$ preserves all information represented in $\bm{W}_U$, contained in $\bm{W}$;
\item $\bm{W'}_U$ is conditionally independent of $T$ given $U$. That happens when $\bm{W'}_U$ is trace-free from $\bW_T$.
\end{enumerate}


The two criteria also allow us to define three scenarios where $\phi$ is not performing perfectly. The first scenario is when $\phi$ is \emph{under-distilling}. That is, $\phi$ produces a representation that satisfies the first criterion but not the second. An under-distilled representation still contains traces of $\bW_T$. An extreme case would be that the original text $\bm{W}$ remains unchanged, yielding $\bm{W'}_U=\bm{W}$. The second scenario is when $\phi$ is \emph{over-distilling}. That occurs when $\phi$ produces a representation that satisfies the second criterion but not the first. An over-distilled text removes all traces of $\bW_T$ but also inadvertently removes information about $\bW_U$. Here, the $F$ index with "$0$" stands for partial faithfulness  (think of it as preserving only half the information in $U$), yielding partial identification, and producing an empty representation. The third scenario is when the model fails on both criteria. The distiller $\phi$ produces a representation that fails to retrieve $\bW_U$ from $\bW$ and still contains traces of $\bm{W}_T$.

\subsection{Causal Identification After Distillation}

The level to which causal identification is satisfied follows both the faithfulness level in $\bW_U$ and the performance level of the distiller. If the faithfulness level of $\bW_U$ is perfect and the distiller produces a perfect representation $\bm{W'}_U$, then causal identification is perfectly obtained, as if $U$ was observed. Denote this ATE for $\textrm{ATE}_{\textrm{F}=1,\textrm{D}=1}$, where the $D$ index with "$1$" stands for perfect distiller and the $F$ index with "$1$" stands for perfect faithfulness. Here, the $F$ index with "$0$" denotes partial faithfulness (think of it as preserving only half the information in $U$), yielding partial identification. Similarly, $D$ index with "$0$" stands for partial distillation (where the distiller collected only half the information about $\bW_U$). We call these \textit{partial} quantities \textit{imperfect}. Although in reality the level of faithfulness and distillation is a matter of degree, here we simplify our argument to two levels only. If the distiller is imperfect and faithfulness is perfect, we write "$0$", like so, $\textrm{ATE}_{1,0}$, omitting the indices for brevity.

Recall the DAG presented in Figure \ref{fig:causalmodel}. Based on it, we can fully identify the $\textrm{ATE}_{\textrm{True}}=\E[Y(1)] - \E[Y(0)]$ by adjusting for $\bm{W'}_U$, instead of $U$, if $\bm{W'}_U$ is a faithful representation of $U$. The full proof is provided in the Appendix.

\begin{align*}
\E[Y(t)] &=  \sum_{x \in \mathcal{X}} \sum_{u \in \mathcal{U}} \E[Y|T=t,X,U] p(X,U) \\
&=  \sum_{x \in \mathcal{X}} \sum_{u \in \mathcal{U}} \E[Y|T=t,X,U] p(X) p(U) \\
&=  \sum_{x \in \mathcal{X}} \sum_{\bm{w'} \in \mathcal{\bm{W'}_U}} \E[Y|T=t,X,U] p(X) p(\bm{W'}_U) \\
\end{align*}

This derivation relies on the assumption that $\bm{W'}_U$ is a sufficient statistic for $U$ with respect to the outcome model---that is, conditioning on $\bm{W'}_U$ captures the same confounding adjustment as conditioning on $U$ itself. This holds when faithfulness is perfect ($F=1$). When faithfulness is imperfect, identification is partial, and the magnitude of residual bias depends on how much information about $U$ is lost in the text representation.

 Consequently, $\textrm{ATE}_{\textrm{True}}=\textrm{ATE}_{1,1}$. In the case of imperfection, we are facing scenarios that yield only partial identification of the true ATE, $\textrm{ATE}$. In the binary encoding of the level of faithfulness and distillation, we have three such scenarios---in the continuous case, there are infinitely many ordered scenarios.

 In the first scenario, the faithfulness level of $\bW_U$ is perfect, but the distiller produces an imperfect representation $\bm{W'}_U$, thus yielding the quantity $\textrm{ATE}_{1,0}$. In the second scenario, the faithfulness level of $\bW_U$ is imperfect in the onset and the distiller produces a perfect representation $\bm{W'}_U$. This combination yields the quantity $\textrm{ATE}_{0,1}$. In this scenario, causal identification is partly obtained with the same faithfulness, as if $\bW_U$ was fully observed, because the distiller is retrieving all the information perfectly. In comparing $\textrm{ATE}_{0,1}$ and $\textrm{ATE}_{1,0}$, we note that these quantities must be equal, assuming that there is no heterogeneity in the functioning of the distiller and the faithfulness level. Under this assumption, a sketch of a proof would be based on the following intuition. The quantity $\textrm{ATE}_{1,0}$ is based on a perfectly faithful $W_U$ but imperfect distiller; and $\textrm{ATE}_{0,1}$ is based on the reversed. If the exact same information about $U$ is either "preserved perfectly in $W_U$ and deteriorated by $\phi$," or "imperfectly in $W_U$ and perfectly retrieved by $\phi$," then the result must be the same. While the information gained and lost occurs in different transformations, these two quantities must still be based on the same information---assuming that there is no heterogeneity in where the distribution $\phi$ or $W_U$ is failing or gaining.

 In the third scenario, the faithfulness level of $\bW_U$ is imperfect, and the distiller is also an imperfect representation $\bm{W'}_U$, then causal identification is partly obtained. That quantity is $\textrm{ATE}_{0,0}$.



\subsection{Key assumptions of distillation}
In practice, we do not observe $U$; neither do we observe all text snippets from $\bW_U$, $\bW_U$, $\bW_R$, nor how $f_W$ weaves together text to produce $\bW$. Thus, although the conditions described define the principles of treatment leakage, it is not possible to directly evaluate whether these conditions are met in many practical cases. Nonetheless, there will likely be a number of ways to indirectly evaluate the magnitude and trace of treatment leakage. For example, $f_W$ is a model of how humans in the research domains of interest produce a full document, using treatment, confounding, and other unrelated text. If scholars have information about this text production, then that provides a way to distill $\bW_U$. Accordingly, we describe a set of assumptions about the relationship between the treatment status $T$ and the text. When some of these assumptions can be justified, scholars may use automatic or semi-automatic distillers.


\paragraph{Separability assumptions.} The approaches we describe below assume in different ways that the effects of $T$ and $U$ can be isolated from each other.
The \textbf{passage separability assumption} states that the document $\bm{W}$ can be partitioned into non-overlapping text portions $\bm{W}_T$, $\bm{W}_U$, and $\bm{W}_R$. This essentially means that the treatment leakage is localized to isolated sections of the text so that if we have prior knowledge about the location or a mechanism to identify $W_T$, we can employ perfect distillation simply by removing these portions.

Passage separability may be too strong in some cases: for instance, if $T$ corresponds to a stylistic or emotional property of the text, such as politeness or sentiment, its effect cannot be isolated from the rest of the text.
In these cases, the \textbf{linear representation separability assumption} may instead be more realistic. This assumption states that $U$ and $T$ correspond to separate linear subspaces in the space of representations $\phi(\bW)$.

For instance, in a representation based on LDA topics \citep{blei2003}, we could assume that $T$ and $U$ are reflected in different topics. In a neural embedding space, the assumption would correspond to the idea that concepts correspond to direction in the representation space \citep{mikolov2013linguistic}, recently referred to as the \emph{linear representation hypothesis} in language model interpretability research \citep{park2024linear}.
If the subspace $\phi_T(\bW)$ corresponding to $T$ can be identified, we can ``scrub'' the representation from the direct causal influence of $T$ by projecting $\phi(\bW)$ into the subspace orthogonal to $\phi_T(\bW)$  \citep{ravfogel2020}. In the simplest case, this corresponds to removing dimensions from the feature representation (e.g., removing treatment-caused topics).

\paragraph{Visibility assumptions.}
If we do not know a priori how $T$ and $U$ affect $\bW$ through $f_W$, it will be hard to isolate the trace and magnitude of treatment leakage since they are associated. Their causal correlation flows through two DAG paths by assumption: $U \to \bW$ and $U \to T \to \bW$.  This flow implies, for instance, that if some aspect of the text is predictive of $U$, it is also indirectly predictive of $T$. Yet, in practice, the strength of these associations will remain challenging to test.
%
%
To handle this practical challenge, we require an assumption of \emph{visibility},
which is the idea that associations arising from direct causal effects of the treatment on the text 
are statistically stronger than the indirect associations between $T$ and $\bW$ via $U$. 
As for separability discussed above, we can make this assumption about the text passages or about the representation space.
%

The \textbf{passage visibility assumption} means visibility at the text passage level so that the passages of $\bW$ that are most strongly predictive of $T$ are those that are directly caused by $T$. This assumption is a stronger version of passage separability defined above: with passage separability, we simply assume that we can separate the text into portions $\bW_T$ and $\bW_U$ that are caused by $U$ and $T$, respectively, while with passage visibility we make the additional statistical assumption that the passages in $\bW_T$ are more strongly predictive of $T$ than those in $\bW_U$.

The second type of visibility assumption is the \textbf{linear representation visibility assumption}. Instead of referring to the text passages, this assumption refers to the level of text representations $\phi(\bW)$. In this case, we assume that the vector space directions that are statistically most strongly associated with $T$ are also those that are caused by $T$.
To exemplify this assumption, if we use an LDA topic model to represent texts, some topics may be strongly correlated with the treatment $T$ since those topics directly describe the treatment.

%


\paragraph{Treatment description assumption.}

In some application scenarios, we may have additional textual information in addition to the documents $\bW$. In particular, the researcher may have access to a set of text snippets that are \emph{known} to be directly related to the treatment status $T$. We refer to these texts as \emph{treatment exemplars}, $\bW_{Tex}$.

Such exemplars are useful in a number of ways. For instance, as discussed later, they can be used to train a supervised model $m(\cdot)$ to detect which text snippets likely belong to $\bW_T$. Thus, we can obtain predictions $\hat{\bW}_T=m(\bW_{Tex})$ which can then supply to a distiller $\phi(\hat{\bW}_T, \bW)$ to negate traces of treatment text in $\bW$, and thereby, producing a higher quality representation $\bm{W'}_U$.

\begin{table}[htb]
\centering
\caption{Mapping of text distillation methods to required assumptions. PSA = Passage Separability Assumption; LRSA = Linear Representation Separability Assumption; PVA = Passage Visibility Assumption; RVA = Representation Visibility Assumption; TDA = Treatment Description Assumption.}
\label{tab:distiller-assumptions}
\begin{tabular}{lcccccc}
\toprule
\textbf{Distiller} & \textbf{Level} & \textbf{PSA} & \textbf{LRSA} & \textbf{PVA} & \textbf{RVA} & \textbf{TDA} \\
\midrule
Human annotator & Passage & \checkmark & & & & \checkmark \\
Similarity-based & Passage & \checkmark & & \checkmark & & \checkmark \\
Distant supervision & Passage & \checkmark & & \checkmark & & \\
Topic removal (LDA) & Representation & & \checkmark & & \checkmark & \\
INLP & Representation & & \checkmark & & \checkmark & \\
\bottomrule
\end{tabular}
\end{table}

\section{Automatic, Semi-automatic, and Manual Distillers}\label{sec:distmethods}

We developed four different automatic or semi-automatic distillers that can be applied for treatment-leakage sensitivity analysis or for entirely removing leakage. As will be shown, these different distillers require different assumptions---some stronger and some weaker---and therefore suitable for a variety of contexts. While the distillers we present are not exhaustive for all kinds of distillers one could imagine, they still cover the majority of cases that applied research can find itself in, wanting to mitigate the contamination of leakage or at least conduct a treatment-leakage sensitivity analysis.

Distillers operate at two stages: the text level or the representation level. \emph{Text-level} methods remove selected passages before computing the document representation. \emph{Representation-level} methods transform an existing representation to eliminate the effect of $\bW_T$. Below, we describe two methods at each level, plus human annotation as a baseline.

All four methods include a stringency hyperparameter that requires careful tuning. Too stringent, and the distiller over-distills: it strips away not only treatment information but also confounder information from $U$, degrading $\bm{W'}_U$. Too lax, and it under-distills: treatment traces from $\bW_T$ persist in the text.

Under the visibility assumption, a distiller can err in two directions. Over-distillation removes $T$'s effect but also partly or fully erases $U$, yielding $\textrm{ATE}_{1,0\uparrow}$ (the ``$\uparrow$'' denotes over-distillation). Under-distillation removes $T$'s effect incompletely, leaving $U$ intact but treatment traces behind, yielding $\textrm{ATE}_{1,0\downarrow}$.

The visibility assumption is key. It states that snippets predictive of $T$ or treatment exemplars $\bW_{Tex}$ are primarily generated by $T$---these snippets belong to $\bW_T$ and cause treatment leakage. The distiller removes them in descending order of treatment-predictability, a process we call \textit{deletion by descension}. Because the optimal stopping threshold remains unknown, distillers can err: stopping too early leaves $\bW_T$ traces (under-distillation); stopping too late erodes $\bW_U$ (over-distillation). The visibility assumption thus provides a principled deletion order. Without it, a distiller lacks a clear starting point and may \textit{ambivalently distill}---removing snippets from both $\bW_T$ and $\bW_U$ indiscriminately.

The visibility assumption will likely hold when the ideal-typical DAG (Figure~\ref{fig:causalmodel}) applies. Critically, treatment-leakage sensitivity analysis requires both text-generating paths: $U \to \bW$ and $T \to \bW$.

Consider what happens when only one path exists. If the leakage path $T \to \bW$ exists but the proxy path $U \to \bW$ does not, distillation reduces post-treatment bias but serves no purpose---$\bW$ contains no information about $U$. Conversely, if $U \to \bW$ exists but $T \to \bW$ does not, distillation backfires. The distiller detects $T$-correlated signals, but these flow through the backdoor path $T \leftarrow U \rightarrow \bW$. Thinking it removes $\bW_T$ information, the distiller actually removes $\bW_U$ information, introducing confounding bias. Domain expertise must establish that both paths exist.

Over- and under-distillation add a second dimension to distiller imperfection. The previous section assumed imperfect distillers would under-distill, preserving the ATE ordering. Under-distillation leaves residual treatment-leakage bias (post-treatment bias). Over-distillation eliminates all $\bW_T$ traces but also erodes $\bW_U$, so remaining bias stems from unobserved confounding rather than post-treatment contamination. A distiller's ``perfection'' reflects how precisely it targets treatment signals; over- versus under-distillation reflects how much it removes.

\subsubsection{Human-annotator distiller}

The most direct approach to text distillation is manual annotation, where human coders read each document and identify passages that contain treatment-related content. This approach requires the \emph{passage separability assumption} (so that treatment content can be localized to specific passages) and implicitly relies on the \emph{treatment description assumption} (coders must understand what treatment-related language looks like).

The workflow proceeds as follows: (1) develop a codebook defining treatment-related content based on domain expertise; (2) train coders on the codebook using example documents; (3) have coders independently flag treatment-related passages in each document; (4) resolve disagreements through discussion or majority voting; (5) remove flagged passages before computing text representations.

While human annotation provides a gold standard against which automatic methods can be validated, it has significant limitations. First, it is costly and time-consuming, particularly for large corpora. Second, coders may introduce inconsistencies, especially when treatment signals are subtle or when treatment and confounding information co-occur within passages. Third, the approach assumes that coders can reliably distinguish content caused by treatment from content merely correlated with treatment---a distinction that may be difficult even for domain experts \citep{audinet2024llms}. For these reasons, we focus primarily on automatic distillation methods, while recognizing that human annotation remains valuable for validation and for small-scale applications.

\subsubsection{Similarity-based Passage Distiller}

The first distillation approach, \emph{similarity-based passage removal}, relies on the \emph{passage separability} and \emph{treatment description} assumptions --- that is, the researcher has access to a set of texts that are \emph{known} to be directly related to the treatment status $T$. We refer to these texts as \emph{treatment exemplars}.

The idea in this distillation approach is to remove passages that are similar to the corpus of treatment exemplar according to some measure of similarity.
We can consider any type of text similarity: in this work, we simply use the cosine similarity of the bag-of-words representations of the texts.
Given a document $\bW$ consisting of subsections (e.g. sentences or paragraphs) $w_i$, and a corpus of treatment text sections $C_T$, we remove a subsection $w_i$ if
\[
\text{cos-sim}(w_i, C_T) > b
\]
where $b$ is a user-defined threshold between 0 and 1.
The user can then control the aggressiveness of the method by adjusting $b$.
The distilled representation $Z$ is then computed by applying any text representation method to the remaining paragraphs.

To recapitulate, we summarize the main steps of this approach.
\begin{enumerate}
    \item Split each document $\bW$ into passages $w_1, \ldots, w_k$.

    \item Let $C_T$ be the concatenation of all treatment descriptions.

    \item Remove all passages $w_i$ where the similarity between $w_i$ and $C_T$ is greater than the threshold $b$.

    \item Apply the text representation function $\phi$ on the remaining passages.
\end{enumerate}

\subsubsection{Distant Supervision for Passage Classification distiller}

The second distillation approach, \emph{distant supervision for passage classification}, relies on the \emph{passage separability} and \emph{passage visibility} assumptions.
In this approach, we build a text classifier that predicts whether a passage (paragraph or sentence) is treatment-related or not. This classifier is applied to the set of passages in each document and the passages that are flagged as treatment-related are removed. Under the PSA, this should be enough for distillation.

However, when training this classifier, there is a technical challenge in that the classifier operates at the \emph{passage} level while the data labeling is available at the \emph{document} level. That is, our dataset tells us for each document what the treatment status $T$ is, but we do not have access to information for a passage about whether it is related to $T$ or not, so
we need a way to bridge the gap between document-level annotation and passage-level decisions.
In principle, we could model the passage-level information as a latent variable \citep{tackstrom2011}, but we opt for a more direct approach inspired by the idea of \emph{distant supervision} \citep{mintz2009}.

We first train a probabilistic classifier at the \emph{document} level and use it to compute a probability of $T=1$ for all \emph{passages} in the dataset. Cross-validation is used when computing these probabilities. Here, by the PVA, we assume that passages where the probability is close to 0 or 1 are those where the treatment-related signal is very strong, and where the probability is close to 0.5, the treatment signal is nonexistent.

In the second step, we train a classifier that acts as a ``tail detector.'' As positive training examples, we select the subset of $N$ passages where the probability is closest to 0 and the $N$ passages where the probability is closest to 1. As negative examples, we sample $2N$ from the remainder of the distribution.
The idea is that this second classifier will distinguish passages expressing a strong treatment signal from those where the treatment status is not obvious. weak signal.

Finally, we apply the second classifier to all passages, and we remove the passages where the probability of a strong treatment signal is greater than a threshold $b$. Again, the user can control the aggressiveness of the method by adjusting $b$.
As above, any text representation method will be used on the remaining paragraphs to compute the distilled representation $Z$.

The following is a step-by-step summary of the approach.
\begin{enumerate}
    \item Split each document $W$ into passages $w_1, \ldots, w_k$.
    \item Using $k$-fold cross-validation, estimate probabilities $\hat{P}(T=1|w_i)$ for all paragraphs $w_i$.
    \item Select the $2 N$ paragraphs in the two tails of the estimated distribution and sample $2 N$ paragraphs from the remainder of the distribution.
    \item Train a classifier $f$ to distinguish the tail paragraphs from the non-tail.
    \item Apply the model $f$ to all paragraphs and remove the paragraphs where the predicted score is greater than a threshold $b$.
    \item Apply the text representation function $\phi$ on the remaining passages.
\end{enumerate}

\subsubsection{Salient Feature distiller}

The third distillation approach, \emph{salient feature removal} relies on the \emph{representation visibility} assumption.
Based on this assumption, we use some measure of statistical association to find the components of a feature representation of a text that are statistically most strongly associated with the treatment variable $T$.
If the assumption holds, distillation can then be carried out by removing the most strongly associated components from the representation.

This approach could be applied to a wide range of language representations. For instance, we could apply it to the raw bag-of-words representation: in this case, we would exclude from the vocabulary the words that are most strongly predictive of $T$.
Previous work has shown that text-based causal inference can be facilitated by applying a more abstract text representation.
In particular, some work has explored the use of representations based on \emph{topic models}, typically based on some variant of latent Dirichlet allocation \citep{blei2003}.
For instance, \citet{roberts2020} used a topic-based text representation in a matching approach, while \citet{sridhar2019} used topics in an approach based on inverse propensity weighting.
In this work, we apply salient feature removal to a topic-based representation.

To quantify the degree of statistical association between a feature and $T$, we can use any measure from the feature selection literature \citep{guyon2003}.
In our experiments, we used the ANOVA $F$-statistic to measure this association. Alternatives include the mutual information and the Gini impurity, commonly used in decision tree learning.


To recapitulate, the approach consists of the following steps:
\begin{enumerate}
    \item Compute topic representations $\phi(\bW)$ for the collection of documents $\bW$.
    \item Using some association measure, rank the topics by their statistical association to the treatment variable $T$.
    \item Compute a distilled representation $Z$ by removing the $K$ most strongly associated topics from the topic representation $\phi(\bW)$.
\end{enumerate}

The user can then control the aggressiveness of the method by selecting the number $K$ of topics to remove from the representation.

\subsubsection{Iterative Nullspace Projection distiller}

The fourth distillation approach, \emph{iterative nullspace projection} (INLP), is a direct application of the method by \citet{ravfogel2020}, which was originally developed for the purpose of learning gender-invariant language representations.
For this method to be applicable in our case, we rely on the \emph{representation visibility} assumption.
In this approach, the idea is (as in the previous method) that we want to remove the treatment-related signal from a previously computed representation $\phi(W)$. In principle, $\phi(W)$ can be a basic bag-of-words representation, a topic-based representation, or a more modern transformer-based approach.

This method operates by finding the linear separator $f$ that best predicts the attribute, and then projects the representations onto a subspace orthogonal to $f$.
In the algorithm described by \citet{ravfogel2020}, INLP is run iteratively for several steps: intuitively, the first steps neutralize the parts of the representations that are most predictive of the attribute.

We applied INLP as a method to remove the $T$-related signal from $\phi(W)$.
However, if we naively apply the INLP method, we risk removing not only treatment-related information, but also information about the unseen confounder $U$ since $T$ and $U$ are correlated.
By the RVA, we assume that the direct effect of $T$ on $\phi(W)$ is stronger than that of $U$, so this would mean that the first iterations of INLP would neutralize $T$ while the effects of $U$ -- which are also predictive of $T$ -- would be neutralized by later iterations.
%
For this reason, the number $N$ of INLP iterations functions as the aggressiveness hyperparameter for this method.

The approach can then be summarized as follows:
\begin{enumerate}
    \item Compute text representations $\phi(W)$ for the collection of documents $W$.
    \item Compute a projection matrix $P_N$ using $N$ iterations of the INLP algorithm.
    \item Compute a distilled representation as the projection $Z = P_N \cdot \phi(W)$.
\end{enumerate}
In this work, we applied INLP to bag-of-words representations of the text documents.

\section{Experimental evaluations of text-distillers}

Validating causal inference methods can often be difficult in practice because the true causal effect may not be known. For this reason, we use two strategies to investigate the distillation methods empirically.
Firstly, following previous work that used synthetic data to validate causal inference methods, we describe how we applied the distillation methods and causal inference methods to synthetically generated texts and covariates. These experiments are described in Subsection~\ref{ss:simulations}.
Secondly, we apply the distillation methods to textual descriptions of International Monetary Fund programs.
We describe these investigations in Subsection~\ref{ss:imfexperiments}.

\subsection{Simulation Design}
\label{ss:simulations}

We carried out a simulation to investigate the behavior of the methods in an idealized scenario where we know the true effect.
We first summarize the general approach in this section and provide details for the simulation in the next section.

\subsubsection{Generating Synthetic Text Data: Text and Numerical Representations}

We generated synthetic data for the simulation by applying ancestral sampling in the causal model presented in Figure~\ref{fig:causalmodel}.
For each instance $i$, we first draw observed and unobserved numerical confounding variables $\bm{X}_i$ and $U_i$, and the treatment status $T_i$ conditioned on $\bm{X}_i$ and $U_i$, and the outcome $Y_i$ conditioned on all previous variables.
The final missing piece is the document $\bm{W}_i$. How can we generate $\bm{W}_i$ conditioned on $U_i$ and $T_i$?

\citet{wooddoughty2021} described an approach to generate synthetic documents for the evaluation of text-based causal inference method by sampling from a GPT-2 model \citep{radford2019}, conditioned on a confounding variable.
In contrast to their approach, text generation is conditioned not only on $U_i$ but also on $T_i$, and in the simulation we need to model how the generation of $\bm{W}_i$ interacts with these two variables.

The approach we selected is designed so that the generated texts satisfy the \emph{passage separability assumption} introduced in Section~\ref{sec:principles}.
%
%
%
We achieve this by conditioning the generation of each paragraph on a randomly selected single paragraph-level topic.
Some topics are associated with $U$, some
with $T$, and some with a residual topic not related to $U$ or $T$. For
each paragraph topic, we define a number of
prompts and a distribution shift that increases the
probability of generating topic-related keywords.

To generate a document $\bm{W}_i$ conditioned on $U_i$ and $T_i$, we generate one paragraph at a time.
For each paragraph $\bm{W}_{ij}$, we draw a paragraph topic $Z_{ij}$ from the set of topics, conditioned on the values of $U_i$ and $T_i$, and a
prompt $W^0_{ij}$ depending on the value of $Z_{ij}$.
Finally, we sample from the GPT-2 language model\footnote{We used the implementation from the HuggingFace repository, \url{https://huggingface.co/gpt2}.} to generate the paragraph text $\bm{W}_{ij}$, starting from the prompt $W^0_{ij}$ and with a vocabulary distribution shift $\theta_{Z_{ij}}$ conditioned on $Z_{ij}$.
Algorithm~\ref{alg:syngen} summarizes the description above in pseudocode for the generation of the numerical values and the documents.

\begin{algorithm}
\caption{Generation of synthetic data.\label{alg:syngen}}
\begin{tabbing}
\hspace{1mm}\textbf{for} $i \in 1, \ldots, N$\\
\hspace{5mm}$\bm{X}_i \sim f_X$\\
\hspace{5mm}$U_i \sim f_U$\\
\hspace{5mm}$T_i \sim \text{Bernoulli}(\text{sigmoid}(f_T(\bm{X}_i, U_i)))$\\
\hspace{5mm}$Y_i \sim f_Y(\bm{X}_i, U_i, T_i)$\\
\hspace{5mm}\textbf{for} $j \in 1, \ldots, K$\\
\hspace{9mm}$Z_{ij} \sim \text{Categorical}(f_{Z}(U_i, T_i))$\\
\hspace{9mm}$W^0_{ij} \sim \text{Categorical}(f_{W^0}(Z_{ij}))$\\
\hspace{9mm}$\bm{W}_{ij} \sim \text{LM}(W^0_{ij}, Z_{ij})$
\end{tabbing}
\end{algorithm}

In the pseudocode above, the functions $f_X$, $f_U$, $f_T$, and $f_Y$ define the distributions of the observed confounders, unobserved confounder, treatment, and outcome, respectively.
On the paragraph level, the function $f_Z$ defines a categorical distribution over paragraph topics, and $f_{W^0}$ a categorical distribution over prompts.

Similar to \citet{wooddoughty2021}, we use two mechanisms to condition the generation of a paragraph on a topic $Z$: a prompt and a vocabulary distribution shift.
The distribution shift is designed to promote a set of \emph{keywords} related to the topic, and we implement it by multiplying the language model probabilities by a topic-specific vector $\theta_Z$ of scale factors:
\[
P'(w|\text{context}, Z) \propto P_{\text{LM}}(w|\text{context}) \cdot \theta_Z(w)
\]

\subsubsection{Generating the Experimental Data for Simulations}
\label{app:specific}

To investigate the variability of estimates, we created 1,000 batches. Each batch consisted of 10,000 instances of the numerical data $T$, $U$, $X$, $Y$ paired with a corresponding document $W$. Each ATE estimate was done using such a 10,000-instance batch.

We used the following distributions to generate the document-level variables: $f_X$ was a 3-dimensional isotropic Gaussian; $f_U$ was an even coin toss; $f_T$ was linear in $\bm{X}_i$ and $U_i$; $f_Y$ was Gaussian with a mean defined by a linear function of $\bm{X}_i$, $U_i$, and $T_i$ and a fixed standard deviation.
Each document consisted of $K$ = 20 paragraphs. For the paragraph generation, we defined five different topics: two corresponding to positive and negative treatment values; two corresponding to positive and negative values of the unobserved confounder; one general background topic that was unrelated to $U$ or $T$ (but conceptually thought of as controlled by other ``residual'' variables $R$).
For a document with given values of $U$ and $T$, we set the topic distribution $f_Z$ to select the $U$ topic with a probability of 0.2, the $T$ topic with a probability of 0.2, and the general topic with a probability of 0.6.

The generated texts were designed to simulate a hypothetical use case where the researchers want to investigate the effect of IMF programs on some country-level indicator \citep[cf.][]{daoud2019international}. The treatment variable $T$ represents the presence or absence of an IMF program; the unseen confounder $U$ represents the political situation of the country with respect to the IMF.
For each topic except the general topic, we define four different prompts: for instance, for a positive treatment value, one of the prompts was \emph{The International Monetary Fund mandates the deregulation of [COUNTRY]'s labor market}. In the analysis, \emph{``[COUNTRY]''} is substituted by randomly sampled country names.

All topics except the general topic defined a distribution shift used when generating from the language model. We used 8 topic keywords for each of these topics. For these keywords, the corresponding entries in the vocabulary distribution shift vector $\log \theta_Z$ were set to a value that defines the strength of the effect of $T$ on $\bW$; for all other words except these keywords, $\log \theta_Z$ was 0.
Since our focus in this paper is on a clear-cut use case where the effects are strong, we set the strength parameter to a value of 4, which gives a noticeable effect on the generated texts.


\subsubsection{Distillation and Estimation Methods}

For each run in the simulations, where we work with a collection of 10,000 documents, we applied all four distillation methods described in Section~\ref{sec:distmethods}. We estimated the average treatment effect (ATE) and compared it to the ground truth value.

The ATE is defined as $\tau = \mathbb{E}[Y_i(1) - Y_i(0)]$, where $Y_i(t)$ is the potential outcome for unit $i$ under treatment $t$. It can be identified in randomized experiments \cite{Rubin1974}. However, the situation is  more complicated in the observational setting, where the treatment is not randomized to units but could be correlated with confounders, $\bX_i$, that are associated with the treatment and the outcome.

To estimate these scores, we applied a $L_2$-regularized logistic regression model using the \texttt{scikit-learn} Python library.\footnote{\url{https://scikit-learn.org}}
When estimating propensities, we represented the (non-distilled or distilled) document as an $L_2$-normalized TF-IDF vector using the 256 most frequent terms in the vocabulary, while the numerical covariates $\bX$ were standardized.

\subsection{Distillation results}
\begin{figure}
    \centering
    \includegraphics[width=1.1\linewidth]{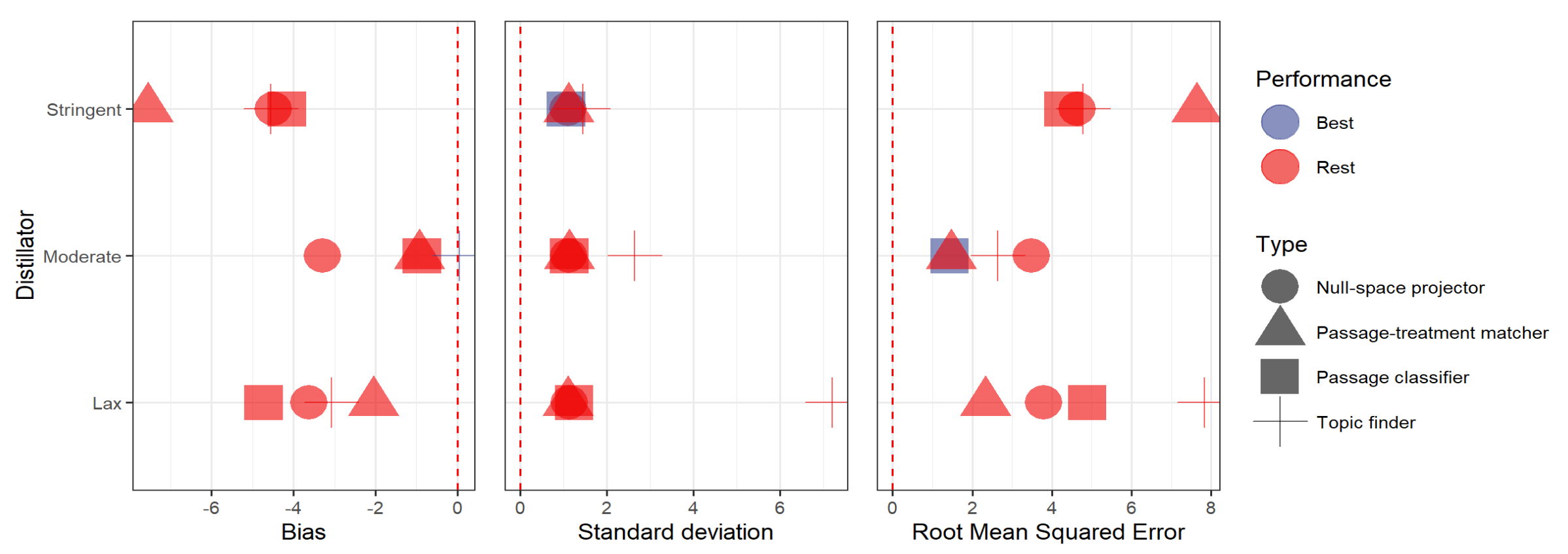}
    \caption{The bias-variance trade-off in text distillation. Over-distillation (high stringency) reduces treatment leakage but removes confounding information, increasing variance and potentially introducing attenuation bias. Under-distillation (low stringency) preserves confounding information but leaves treatment leakage, inducing post-treatment bias. Optimal distillation balances these competing errors.}
    \label{fig:distillers}
\end{figure}

Figure \ref{fig:distillers} shows the performance of our distillers. In terms of bias, moderate distillation generally works better than either stringent or lax. As previously discussed, the results align with our intuition because removing too much information about the treatment also removes information about the confounder. Removing too few leaves leaves traces of treatment leakage. In both cases, the distillers produce biased results. For the moderate distillers, the topic finder reduced bias almost perfectly, yielding a result close to zero. The two close runners-up are the passage-treatment matcher and the passage classifiers, which produce nearly identical results.  The null-space project lags far behind the other distillers.

For variance (Figure \ref{fig:distillers} shows standard deviation to align scales with bias), the topic finder underperforms relative to other distillers---not dramatically, but noticeably. The moderate topic finder's standard deviation reaches 2.2, compared to 0.5 for other methods. The topic finder also shows stark variability between lax and stringent settings, with a difference of 6 standard deviation points. Other distillers show comparable variability across stringency levels. The stringent passage classifier achieves the lowest variance, though by a small margin.

Figure \ref{fig:distillers} also displays root mean squared error (bias squared plus variance) for completeness.

\section{Application: Impact of the International Monetary Fund (IMF) Policy on Child Health}
\label{ss:imfexperiments}
While much research has been devoted to evaluating the policy effects of International Monetary Fund (IMF) programs on vulnerable populations \citep{stuckler2008}, such as children \citep{daoudImpactInternationalMonetary2017}, scholars have yet to arrive at a consensus about the magnitude, let alone the direction, of such policies \citep{stubbs2020}. The IMF is an international organization with a mission to assist countries in macroeconomic crises, but it is also known for its austerity policies \citep{babb2005}. When governments ask for IMF assistance, the IMF evaluates that country's macroeconomic conditions, what macroeconomic treatment is appropriate, and an assessment of the government's political willingness to successfully implement often stringent neoliberal policies. A government's political willingness relies not only on the implementation difficulty of the treatment (e.g., privatizing state-owned companies vs.\ lowering fiscal spending) but also on the country context. Poorer countries will be inclined to endure difficult treatments, and thus, have a higher political will to select into IMF programs than richer countries \citep{vreeland2003}.

Determining the impact of IMF programs poses a fundamental causal identification challenge \citep{dreher2009,imbens2015}. Consider an analogy: if we studied doctor visits by comparing healthy and unhealthy people who see doctors, we might conclude that doctor visits cause poor health. Identifying the true causal effect requires comparing what would have happened to unhealthy patients without care against what happens when they receive care. The difference, aggregated across a representative sample, yields the causal effect. The IMF faces the same challenge: governments are the patients, and the IMF is the doctor.

The key challenge is assessing \textit{why} governments select into IMF programs and their \textit{political will} to endure stringent austerity policies. Macroeconomic conditions---encoded in tabular country-level data $\bX$---reveal part of the story. But political will remains obscure, unobserved in existing datasets.

Researchers have addressed unobserved political will through Heckman selection adjustment, instrumental variables, and covariate adjustment \citep{stubbs2020}. The most credible approaches rely on instrumental variables, yet instruments remain contested \citep{imbens2015,deaton2010}. Even valid instruments shift the estimand from the Average Treatment Effect (ATE) to the Local Average Treatment Effect (LATE), which captures only a narrow slice of the variation, limiting generalizability \citep{imbens1994}.

If political will could be measured systematically, scholars could estimate IMF policy effects more credibly than existing strategies allow \citep{dreher2009}.

IMF officials assess political will before programs begin, and these assessments are digitized and publicly available in IMF policy archives. Among several document types, IMF Executive Board Specials (EBS) documents prove particularly valuable. EBS documents formalize the Executive Board's decision-making process---the body that directs IMF funding. IMF officials produce these steering documents for each Board meeting, and the Board relies on them when deciding whether to engage a government and provide resources.

An EBS describes macroeconomic issues, political-economic contexts, policy conditions, and each government's track record with the IMF---including its likelihood of complying with IMF policies. The documents evaluate governments' political motivation and past performance implementing IMF programs \citep{vreeland2003}. Scholars can therefore measure political will by processing EBS documents for text-based causal inference \citep{roberts2020,mozer2020}, building on NLP methods that extract policy conditions from IMF loan agreements \citep{daoudNLPpolicy2019}. This approach targets unmeasured confounding more directly than existing identification strategies.

Why have scholars not used this strategy? First, text-based causal inference methods have only recently emerged. Second, human annotation is costly: EBS documents often exceed 80 pages, requiring thousands of hours to annotate and maintain.

Third---and most relevant here---EBS documents risk treatment leakage. They describe not only political will but also the treatment itself: the IMF policies governments will implement \citep{daoud2019international}. This leakage requires treatment-leakage sensitivity analysis to bound the estimated effect. The analysis yields maximum and minimum IMF effects across distillation levels, covering the range a leak-free analysis would produce.

Our analysis pursues two goals. First, we compare estimates with and without EBS adjustment (always including tabular controls) to determine how much additional confounding information EBS documents capture beyond country-level data.

Second, we apply treatment-leakage sensitivity analysis using our distillers. Starting with full leniency and increasing stringency, we re-estimate IMF effects at each level. Although we cannot estimate leakage magnitude without ground truth, oscillating estimates---as seen in simulations---signal leakage traces.

Leakage and confounding information are deeply intertwined, creating a trade-off. Aggressive distillation removes leakage (reducing post-treatment bias) but also strips confounding information (increasing confounding bias). Lenient distillation preserves confounding information but leaves leakage traces. Our sensitivity analysis bounds causal effects across this trade-off, from most lenient to most stringent.

\subsection{Causal identification and estimation strategies}

\begin{figure}[htb]
    \begin{center}
\subfloat{}{
    \begin{tikzpicture}[line width=0.25mm, scale=1.0, transform shape]
        \node (X) {$\textbf{Country context}$} ; %
        \node[below=of X, text=gray] (U) {$\textbf{Political will}$} ; %
        \node[right=of X, xshift=-6mm, yshift=-8.5mm] (T) {$\textbf{IMF policy}$} ; %
        \node[right=of T, xshift=-2mm] (Y) {$\textbf{Child Health}$} ; %
        \node[right=of Y, xshift=-3mm, yshift=0mm, text=gray] (R) {$R$} ; %
        \node[right=of U, xshift=9.5mm] (W) {$\textbf{EBS document}$} ; %
        \edge[-latex] {X,U} {T} ; %
        \edge[-latex] {X,U,T} {Y} ; %
        \edge[-latex] {U} {W} ; %
        \edge[-latex] {R} {W} ; %
        \edge[-latex, line width=0.33mm, red] {T} {W} ; %
    \end{tikzpicture}
}
\end{center}
\caption{A causal model consisting of observed variables: observed confounders (country context, $\bX$), IMF policy (treatment, $T$), child health (outcome, $Y$), and  IMF's EBS documents ($\bW$). The main unobserved confounding (shown in grey) is political will ($U$) and residual factors ($R$). As before, the red-colored edge represents treatment leakage. Our distillation-sensitivity test consists of a set of function $f$ that estimates the treatment information in the text and removes it, leaving only information from the unobserved confounder, political will. A perfect distiller, $f$, is a procedure that finds all leakage and, thus, removes all information flowing via the red arrow. When the distiller is applied to ($\bW$), that is equivalent to deleting the red arrow; a less-than-perfect intervention reduces at least the strength of the red arrow. \label{fig:IMFcausalmodel}}
\end{figure}
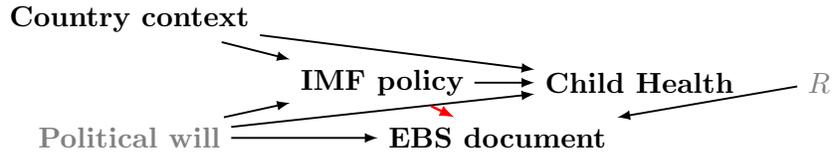

Let us clarify our identification argument formally. Figure \ref{fig:IMFcausalmodel} illustrates our identification strategy in a DAG. The outcomes of interest are child health, operationalized as the proportion of children that receive four types of essential immunizations (Tuberculosis, Diphtheria, Meningococcus, and Polio), and the proportion of children that die before the age of five. We selected five outcomes to receive a wider picture of the impact of the IMF on children than merely selecting one outcome. The treatment is public-sector employment policy, defined as those countries that have at least one such policy $T=1$ backed in to their IMF program; the control $T=0$ is those countries with zero such policies bundled into their IMF program. IMF public-sector employment policies have been shown to carry a detrimental effect on children, because they adversely affect the number of employed in the health care system \citep{daoudReinsberg2018,daoudImpactInternationalMonetary2017}.

For an exhaustive substantive study, future research may analyze the impact of any combination of IMF policies.

Because EBS documents exist mainly for those countries that enroll in IMF programs, our causal estimand is then the average treatment effect on the treated (ATT), written as the difference in potential outcomes $\textrm{ATT}=\E[Y(1) - Y(0)|T=1]$. As shown in our DAG, this estimand is identified by including the EBS document ($\bW$) as a proxy for political will and other country context factors $\bX$, as these will block all backdoor paths. These $\bX$ consist of a set of factors normally included in IMF research, presented in the Appendix.

As previously discussed, an EBS document ($\bW$), contains information about unobserved political will ($U$), and thus there is an arrow between them. But an EBS document ($\bW$), contains information not only about political will ($U$) but often also about IMF policy intervention that a government has to implement ($T$). This leakage is represented by the red arrow in Figure \ref{fig:IMFcausalmodel}. The larger the leakage, the more post-treatment bias there will be when adjusting for ($\bW$). Although a point estimate of the magnitude of leakage will remain unknown -- for that, we will need some information about how exactly political will manifest in a sample of cases -- our proposed methods will evaluate a range of potential magnitudes. This range provides the bounds for our causal estimates.

Lastly, EBS documents also contain other snippets of text irrelevant to either $T$ and $U$; call them residual text $R$. These $R$ induce essentially statistical noise that adversely affects the variance of our causal estimator. If $R$ is extensive, it may nullify the value of including text $\bW$ in the adjustment set.

For causal estimation, we use the same procedure as in the simulation study. First, we estimate the propensity of selecting into IMF programs in two ways. One where we do not include text $\bW$, $\pi(\bX)=P(T=1|\bX)$, and one where we do, $\pi(\bX,\phi(\bW))=P_{\text{LM}}(T=1|X,\phi(\bW))$. Here $\phi(\bW)$ is a TDM text representation (16,384 terms) with no distillation. As stated before, comparing these two propensities will reveal the difference of including EBS documents as a proxy for political will. For estimating these propensities, we use LASSO models.

Second, we calculate the IPW-weighted mean difference for both propensities \citep{rosenbaum1983,funk2011doubly}. The estimated ATT using text is then given by,
\[
\widehat{\tau} =  \frac{1}{n}\sum_{i=1}^n \left\{ \frac{T_i Y_i}{\hat\pi(\bX_i,\phi(\bW_i))} -\frac{(1-T_i)Y_i}{1-\hat\pi(\bX_i,\phi(\bW_i))}\right\}.
\]
Note that while we present the standard IPW estimator for the Average Treatment Effect (ATE), our empirical application focuses on the Average Treatment Effect on the Treated (ATT). For the ATT, we weight untreated observations by $\hat{\pi}_i/(1-\hat{\pi}_i)$ to create a pseudo-population comparable to the treated group. In both cases, propensity scores near 0 or 1 can produce extreme weights; we address this by using $L_1$-regularized logistic regression, which naturally shrinks extreme predictions, though we acknowledge that formal trimming or truncation could provide additional robustness \citep{crump2009dealing}.

The ATT without using text is estimated equivalently, but leaving out $\phi(\bW_i)$ from the equation. We use 1000 bootstrap samples to estimate the standard error for the ATTs.

Third, for the treatment-leakage sensitivity analysis, we distill $\theta_{k,t,s}(\bW)$ across our four distillers, varying the size of the TDM representation $t \in \{64, 256, 1024, 4096, 16384\}$ $k \in \{1,..,4\}$, and with increasing level of distillation stringency $s \in \{\textit{"most lenient"},..,\textit{``most stringent''}\}$. Larger TDM representation will capture the underlying text patterns better, but is computationally more expensive. The stringency parameter definition varies with the distillation method: passage-treatment matcher (cosine similarity) has $s \in \{0.0001, 0.0002, 0.001, ...,0.1\}$ measuring the vector angle at which a passage points in the same direction as treatment text;  paragraph classifier has $s \in \{0.01, 0.1, 0.2,...,0.999\}$, capturing the probability threshold at which text passages resemble treatment class;  Null-space projector has $s \in \{1,2,3,...,7\}$, indicating the number of projection iterations where each iteration removes the most treatment-predictive linear direction from the representation; and topic finder $s \in \{0,1,2,...,16\}$, designating the number of likely treatment-related topics removed. In total, we run 291 different text representation varying $k$, $t$, and $s$, and thus producing this number of $\widehat{\tau}_{k,t,s}$. All these runs are also bootstrapped at 1000 draws to estimate standard errors.

Our Inverse Propensity Weighting uses \(L_1\)-regularized logistic regression model implemented through the \texttt{glmnet} package in \texttt{R}. The regularization parameter (\(\lambda\)) was automatically selected using 10-fold cross-validation. During the propensity score estimation, we represented the (distilled or non-distilled) document as an \(L_2\)-normalized TF-IDF vector consisting of the 256 most frequent terms in the vocabulary, while the numerical covariates \(\bX\) were standardized.

\subsection{The causal effect of IMF on child health}

\begin{figure}
    \centering
\includegraphics[width=1\linewidth]{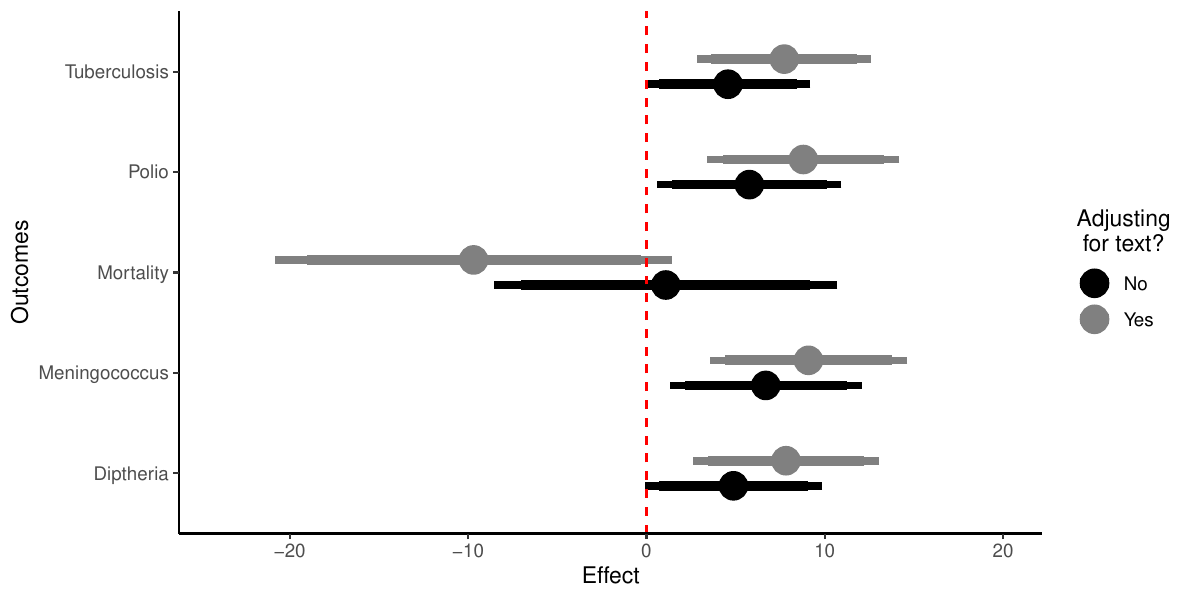}
    \caption{Impact of International Monetary Fund public-sector employment policy on child health}
    \label{fig:IMFimpact}
\end{figure}

Figure \ref{fig:IMFimpact} shows the impact of IMF public-sector employment policy on child health. It compares two sets of estimates with and without text adjustment, adjusting for the same tabular data in both sets. Looking at the tabular-only results (black points), we would conclude that IMF policy tends to increase the percentage of children vaccinated: Polio shows an effect of approximately 7 percentage points (pp) and Meningococcus approximately 6 pp, both significant at the 95\% level; Tuberculosis and Diphtheria show effects of approximately 5 pp each, but only significant at the 90\% level. Under-five mortality is not statistically different from zero in the tabular-only specification (point estimate near 0 pp).

However, adjusting for EBS text (gray points) makes a substantive interpretative difference. All four vaccination outcome estimates shift rightward, indicating larger positive effects: Tuberculosis increases from approximately 5 pp to 8 pp, Diphtheria from 5 pp to 7 pp, Meningococcus from 6 pp to 10 pp, and Polio from 7 pp to 10 pp. Critically, all four vaccination estimates become significant at the 95\% level when adjusting for text. These shifts suggest that tabular-only adjustment leaves residual confounding that attenuates treatment effects, and that EBS text captures additional confounding information---likely reflecting political will---that, when controlled for, reveals stronger positive effects on child immunization.

For under-five mortality, the pattern differs: text adjustment shifts the point estimate from near zero (tabular-only) to approximately $-10$ pp, suggesting a mortality-reducing effect of IMF public-sector employment policies, though this estimate is only significant at the 90\% level. The larger confidence interval for mortality relative to vaccination outcomes reflects greater uncertainty, possibly because mortality is influenced by a broader set of factors beyond those captured by the EBS documents.

Using text for adjustment clearly changes the result enough to produce a different analysis, interpretation, and conclusion. But how susceptible are these results to treatment leakage? The next section addresses this concern.

\subsection{A treatment-leakage sensitivity analysis}

\begin{figure}
    \centering
\includegraphics[width=1.1\linewidth]{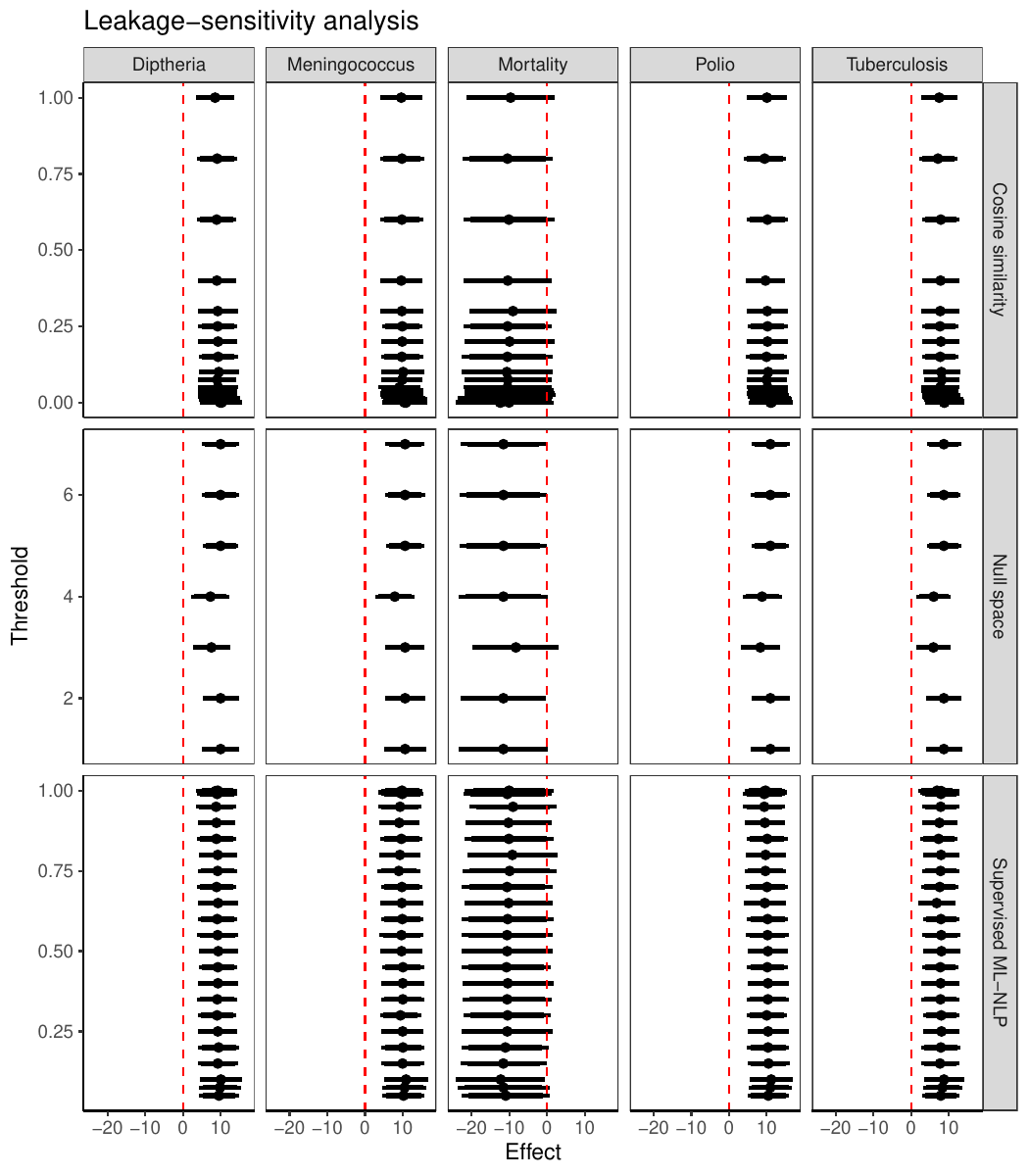}
    \caption{Treatment-leakage-sensitivity analysis: Four distillers searching and cleaning for treatment leakage across the five outcomes. The analysis uses the largest Term-Document Matrix (16 384) for maximum representation. The standard errors use 1000 bootstraps, for each threshold level.}
    \label{fig:sensitivity_analysis}
\end{figure}

Figure \ref{fig:sensitivity_analysis} presents our results from the treatment-leakage sensitivity analysis across three distillation methods (cosine similarity, null-space projection, and supervised ML-NLP) and five child health outcomes. The pattern of leakage would manifest as it did in the simulation: point estimates tracing a convex curve that bends toward the y-axis as distillation stringency increases from lenient (threshold near 0) to aggressive (threshold near 1.0 for similarity-based methods, or higher iteration counts for null-space projection). When leakage is pronounced, this ``belly'' grows larger at moderate distillation levels.

In the IMF case, the sensitivity analysis reveals only minor traces of leakage. For all four vaccination outcomes---Diphtheria, Meningococcus, Polio, and Tuberculosis---point estimates remain remarkably stable in the range of 5--10 pp across all distillation thresholds and methods. The cosine similarity distiller shows estimates clustering tightly around 5--8 pp regardless of threshold, suggesting minimal treatment information in the text representations. The supervised ML-NLP distiller exhibits somewhat more variability, with estimates ranging from approximately 3 pp to 10 pp across thresholds, but without the pronounced convex curvature that would indicate substantial leakage. The null-space projector shows modest movement at threshold level 3, but the curvature is small (estimates shifting by only 1--2 pp).

For mortality, the pattern is noisier: point estimates range from approximately $-15$ pp to $-5$ pp across distillers and thresholds, with wider confidence intervals throughout. This instability may reflect either weak leakage signals that distillers struggle to isolate or greater inherent uncertainty in the mortality outcome, rather than systematic treatment leakage.

\begin{figure}
    \centering
\includegraphics[width=0.9\linewidth]{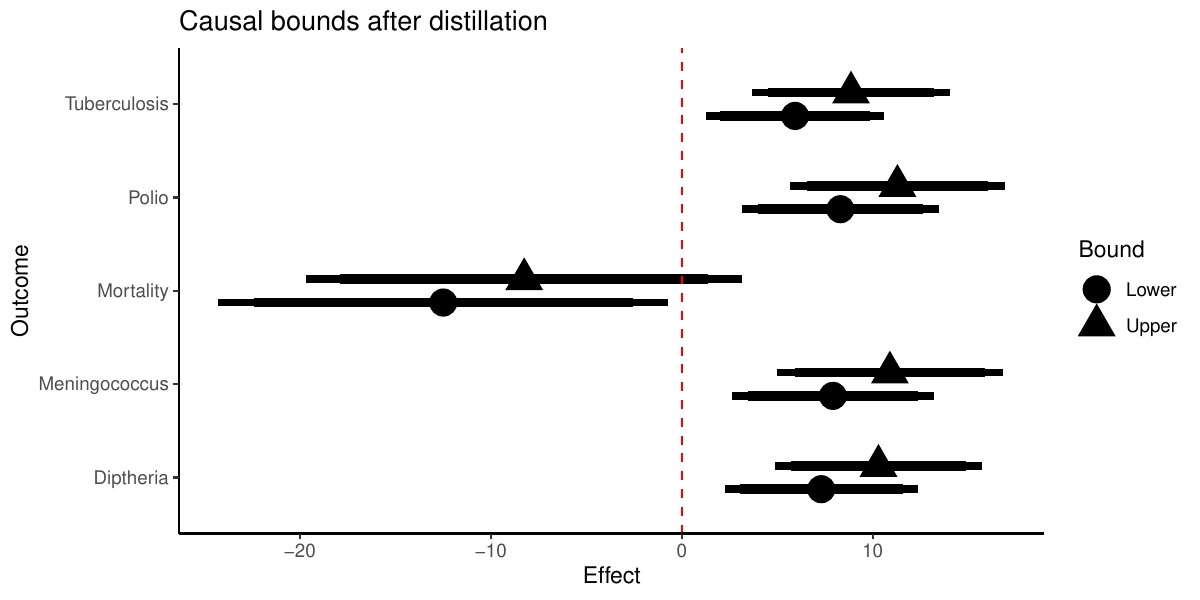}
    \caption{The upper and lower limits show the maximum and minimum effect, respectively, using the treatment-leakage-sensitivity analysis based on a term-document matrix of 16,384. The standard errors use 1000 bootstraps.}
    \label{fig:Bounds}
\end{figure}

To address these doubts, we calculate two bounds for each outcome, shown in Figure \ref{fig:Bounds}. These bounds represent the lowest and highest point estimates found across all distillers and stringency levels in the treatment-leakage sensitivity analysis. For vaccination outcomes, the bounds are reassuringly tight and consistently positive: Diphtheria ranges from approximately 5 pp (lower) to 10 pp (upper), Meningococcus from 5 pp to 12 pp, Polio from 7 pp to 13 pp, and Tuberculosis from 5 pp to 12 pp. Importantly, even the lower bounds remain statistically distinguishable from zero at conventional significance levels, providing robust evidence that IMF public-sector employment policies increase child immunization rates regardless of assumptions about treatment leakage.

For mortality, the bounds span a wider and more uncertain range: the lower bound is approximately $-15$ pp and the upper bound approximately $-5$ pp. While both bounds suggest a mortality-reducing effect, the confidence intervals for these bounds overlap with zero, particularly for the upper bound. This means we cannot verify that the mortality effect is statistically different from the null at the 90\% level when accounting for potential leakage.

Figure \ref{fig:sensitivity_analysis} shows our results from the largest TDM space. A large TDM enables a better representation of the underlying human meaning in text \citep{gentzkow2019text}. As a robustness, we ran our distillation-sensitivity test using the smaller TDMs to evaluate how much that choice may affect the results--shown in the Appendix. Although the baseline estimates (most lenient) tend to start at a lower causal effect from smaller TDS, we see that the trend of distillation is relatively stable across TDM size, which is evidence that our distillation is robust to TDM variation.

\section{Discussion}\label{sec:discussion}

Text-based causal inference often treats documents as proxies for otherwise unmeasured confounders. Our results show that this strategy is fragile when the same documents also encode treatment status. Because language can refer to planned or future actions, treatment leakage can arise even when texts are produced ``before'' assignment in calendar time. The practical implication is that researchers must assess whether a text source behaves like pre-treatment information about $U$ or like a post-treatment collider influenced by $T$.

\paragraph{Key contributions.} First, we clarify treatment leakage within a causal-graphical framework and formalize distillation as surgery on the text channel: construct a representation $\phi(\bW)$ that preserves confounding information while removing treatment-induced variation. This framing reveals why ``collecting text pre-treatment'' does not prevent post-treatment bias \citep{Pearl2015}---what matters is whether the \emph{content} lies causally downstream of $T$.

Second, we provide a practical toolbox of text distillers, each operationalizing different assumptions about how leakage appears. Two passage-level approaches assume separability/visibility: (i) a similarity-based distiller removes passages resembling known treatment exemplars; (ii) a distant-supervision classifier flags treatment-predictive passages using document-level labels. Two representation-level approaches assume representation visibility: (iii) salient feature removal (via topic removal); (iv) iterative nullspace projection (INLP) \citep{ravfogel2020}. All methods expose the core trade-off through a stringency hyperparameter: under-distillation leaves leakage (post-treatment bias); over-distillation strips confounding signals (confounding bias).

Third, we propose treatment-leakage sensitivity analysis as a workflow for settings where leakage is suspected but unmeasured. By re-estimating the causal effect across a spectrum of stringencies and distillers, researchers can diagnose instability patterns consistent with leakage and report bounded effects rather than a single potentially contaminated point estimate. In simulations with known ground truth, moderate distillation generally improved performance relative to lax or overly stringent settings, and distillers differed in bias--variance profiles (e.g., topic-based removal reduced bias but sometimes increased variance). In the IMF application, adding EBS text to adjustment materially changed substantive conclusions relative to tabular-only controls, and the sensitivity analysis suggested only limited traces of leakage; the resulting bounds continued to support positive effects on vaccination outcomes while leaving mortality effects more uncertain.

\paragraph{Limitations of the simulation study.} Our simulation, designed for manipulable and measurable leakage, is idealized. GPT-2 generated paragraphs satisfy passage separability by construction; real documents may intertwine treatment and confounding within passages, express them indirectly, or encode them through style rather than keywords. The data-generating process was simple (binary treatment, linear/logistic relationships), and we used a specific estimation stack (TF-IDF with regularized propensity models). Richer representations, alternative estimators, smaller samples, multi-valued treatments, or time-varying confounding may interact differently with leakage. The simulated signal was strong and structured; weaker or heterogeneous leakage may evade detection and require different diagnostics.

A key limitation concerns the visibility assumptions. These assume that text features most predictive of treatment are directly caused by treatment, not predictive through confounding pathways. When treatment and confounding correlate highly in representation space, aggressive distillation may strip confounding information alongside treatment signals. We cannot verify visibility empirically; researchers must judge plausibility from domain knowledge about how treatment and confounding manifest in their texts. Diagnostics for assumption violations would strengthen these methods substantially.

\paragraph{Future research directions.} First, develop formal sensitivity models linking observable diagnostics to interpretable leakage parameters, enabling uncertainty statements beyond empirical bounds. Second, integrate distillation with estimation: rather than preprocessing and then estimating, co-design representations and estimators that explicitly trade off treatment invariance against confounding preservation---for example, combining distillation objectives with doubly robust estimation.

Third, build more realistic benchmarks: simulations where $\bW_T$ and $\bW_U$ are inseparable, treatment signals are indirect or domain-specific, and document availability is itself selective. This matters as machine learning increasingly taps alternative data sources---satellite imagery for poverty measurement \citep{daoudPovertyIndia2022}, for instance---where similar leakage concerns arise. Curated empirical datasets with passage-level leakage annotations (created via active learning with human or LLM assistance) would sharpen distiller evaluation and stringency guidance. The IMF application suggests substantive extensions: leveraging additional pre-treatment text to move from ATT toward ATE, and studying policy bundles where leakage and confounding vary across document genres.

\paragraph{Practical guidance for applied researchers.}
When should researchers use treatment-leakage sensitivity analysis? Whenever text proxies for unobserved confounding and might also encode treatment information. Examples: policy documents from implementing agencies, administrative records created during treatment, and any text where authors reference future interventions or anticipate treatment assignment.

For distiller selection: with treatment exemplar documents (explicit policy statements), use \emph{similarity-based} or \emph{distant supervision} distillers. Without exemplars, \emph{topic removal} or \emph{INLP} identifies treatment-predictive signals from data alone. When passage separability holds (treatment information is localized), prefer passage-level distillers; when treatment signals pervade documents, representation-level distillers fit better.

For stringency: report bounds across multiple levels rather than choosing one ``optimal'' threshold. Stable estimates across stringency values suggest minimal leakage or entanglement too deep for distillation to untangle. Instability signals sensitivity to distillation choices and warrants interpretive caution.

\paragraph{Broader implications.} ``Text as a proxy'' is a design and validation problem, not a plug-and-play fix for unmeasured confounding. Document why text should encode $U$, articulate how it might encode $T$, and report robustness checks varying leakage removal. Treatment-leakage sensitivity analysis communicates this uncertainty transparently.

As computational social science taps administrative narratives, policy documents, satellite imagery, and LLM-generated summaries \citep{sakamotoEarthObs2024,daoudSatelliteML2024,daoudSatelliteIndia2023,petterssonSatelliteGNN2025}, inadvertent conditioning on post-treatment information grows likelier. Leakage detection and mitigation tools improve credibility and comparability across studies using different text sources. Stable estimates across distillers and stringency levels strengthen confidence; instability is itself informative and should temper policy inference.

\section*{Acknowledgments}
Adel Daoud thanks the Swedish Research Council and WASP-HS for project support.

\bibliographystyle{plainnat}
\bibliography{bibliography}

\section{Appendix}

\subsection{Processing IMF policy documents}
IMF Executive Board Specials (EBS) documents are distributed as PDF files. We converted each PDF to raw text using \texttt{pdfminer.six}\footnote{\url{https://github.com/pdfminer/pdfminer.six}} and applied the cleaning pipeline described by \citet{reichl2021confounder}. Specifically, we removed boilerplate (headers, footers, page numbers, and agenda metadata), joined words split across line breaks, normalized whitespace, and dropped table-like blocks dominated by numbers and punctuation. We then segmented documents into paragraphs using blank lines and punctuation cues, retaining original order for passage-based distillers, and discarding extremely short fragments and duplicates introduced by scanning artifacts. For bag-of-words representations, we lowercased, removed punctuation, and kept alphabetic tokens. We constructed term--document matrices by retaining the $t$ most frequent terms (with $t\in\{64,256,1024,4096,16384\}$) and transformed raw counts to TF--IDF, followed by $L_2$ normalization. These preprocessed representations serve as the baseline (no distillation) adjustment and as inputs to the representation-level distillers.

\subsection{Country covariates}

To adjust for observed confounding, we include a standard set of pre-treatment country covariates $\bX$ widely used in cross-national studies of IMF program selection and health outcomes. These covariates capture (i) macroeconomic need and crisis severity: GDP per capita, GDP growth, inflation, international reserves, current-account balance, fiscal balance, and external debt/debt service; (ii) exposure to the global economy: trade openness and capital account restrictions; (iii) demographic structure relevant for child health: population size, urbanization, and the share of children; and (iv) political and institutional constraints that shape both program participation and implementation capacity: regime type/democracy, executive constraints, conflict, and government turnover. All covariates are measured prior to program approval (lagged by one year) to avoid post-treatment contamination and are standardized before estimating propensities. This specification follows common practice in the IMF causal literature and is used in both the tabular-only and tabular+text propensity models.

\subsection{Treatment-leakage-sensitivity analysis}

The full treatment-leakage sensitivity analysis results across all four distillers, five TDM sizes, and multiple stringency levels are available in the online supplementary materials. Here we summarize the key patterns.

Across all distillers and outcomes, we observe that moderate stringency levels generally produce the most stable estimates. At very low stringency (minimal distillation), estimates resemble those from the baseline text-adjusted model, suggesting that little treatment leakage is being removed. At very high stringency (aggressive distillation), estimates become more variable and often attenuate toward zero, consistent with over-distillation removing confounding information alongside treatment signals.

The passage-level distillers (similarity-based and distant supervision) show more sensitivity to the choice of stringency parameter than the representation-level distillers (topic removal and INLP). This may reflect the discrete nature of passage removal compared to the continuous transformations applied by representation-level methods. The INLP distiller shows particularly smooth degradation as the number of iterations increases, consistent with its design as an iterative refinement procedure.

For detailed figures showing point estimates and confidence intervals across all specifications, we refer readers to the supplementary materials.

\end{document}